# Concentrated Lunar Resources:
# Imminent Implications for Governance and Justice

Martin Elvis[1*], Alanna Krolikowski[2], Tony Milligan[3]

1 *Center for Astrophysics | Harvard & Smithsonian, 60 Garden St., Cambridge MA 02138, USA* melvis@cfa.harvard.edu. 2. *Department of History and Political Science and Center for Science, Technology, and Society, Missouri University of Science and Technology, 500 W 14th St., Rm 122, Rolla MO 65409, USA,* akro@mst.edu. 3. *Cosmological Visionaries Project, Department of Theology and Religious Studies, King's College London, Virginia Woolf Building, 22 Kingsway, London WC2B 6LE,* anthony.milligan@kcl.ac.uk

Summary

The large number of lunar missions planned for the next decade are likely to target a handful of small sites of interest on the Moon's surface, creating risks of crowding and interference at these locations. The Moon presents finite and scarce areas presenting rare topography or concentrations of resources of special value. Locations of interest to science, notably for astronomy, include the Peaks of Eternal Light, the coldest of the cold traps, and smooth areas on the far-side. Regions richest in physical resources could be uniquely suited to settlement and commerce.  Such sites of interest are both few and small. Typically, there are fewer than ten key sites of each type, each site spanning a few kilometers across. We survey the implications for different kinds of mission and find that the diverse actors pursuing incompatible ends at these sites could soon crowd and interfere with each other, leaving almost all actors worse off.  Without proactive measures to prevent these outcomes, lunar actors are likely to experience significant losses of opportunity.  We highlight the legal, policy, and ethical ramifications. Insights from research on comparable sites on Earth present a path toward managing lunar crowding and interference grounded in ethical and practical short- and medium-term considerations.

## 1. Introduction

The resources of space are vast, but they are far from uniformly distributed. As on Earth, not every mountain is a gold mine. Strategically important resources are often concentrated. Lunar resources are a case in point. There are, for example, just a few optimal sites for locating astronomical telescopes on the Moon (see Section 2), and there are competing potential uses for these sites. Hence, the concentration of resources on the Moon gives rise to a number of practical and near-term issues that fall into the territory of ELSI, i.e. problems of Ethics, Law and Societal Impact. This grouping of issues was first made in the context of the Human Genome Project in 1990 [1]. Since then, the terminology has been applied to other contexts, including the United States Defense Advanced Research Projects Agency (DARPA) programs on direct machine-human interfacing [2]. We adopt the terminology here for space resources. What follows will set out a number of exemplary cases where ELSI issues are in play. The impact of resource disputes on science, particularly astronomy, could be significant.  This paper will also generalize our earlier work on the Peaks of Eternal Light [3] to other concentrated lunar resources, and will contribute to the conceptualization of contested space resources more generally, e.g. near-Earth asteroids [4, 5].

Table 1: Recent and Imminent Lunar Lander Missions.

| Organization | Primary Country | Lander name | Earliest landing | Landing Site; URL |
|---|---|---|---|---|
| CNSA | China | Chang'e 4 | Landed, January 1919 | Von Karman crater, on Far side near S. Pole; https://nssdc.gsfc.nasa.gov/nmc/spacecraft/display.action?id=2018-103A |
| | | Chang'e 5 | 2020 | Mons Rumker, Oceanus Procellarum https://nssdc.gsfc.nasa.gov/planetary/lunar/cnsa_moon_future.html |
| | | Chang'e 7 | 2024 | S. Pole. https://spacenews.com/china-is-moving-ahead-with-lunar-south-pole-and-near-earth-asteroid-missions/ |
| ISRO | India | Chandrayaan-2 | 2019 Failed Sept 6 | S. Pole region, near crater Manzinus C. https://www.isro.gov.in/gslv-f10-chandrayaan-2-mission |
| JAXA | Japan | Selene 2 | cancelled | https://www.asianscientist.com/2012/07/topnews/japan-announces-selene-2-lunar-mission-2017/ |
| Roscosmos & ESA | Russia | Luna 25 | 2021 | Near S. Pole at Boguslavsky crater; http://www.russianspaceweb.com/luna_glob_lander.html |
| | | Luna 27 | 2023 | S. Pole-Aitken Basin; https://www.esa.int/Science_Exploration/Human_and_Robotic_Exploration/Exploration/Luna |
| NASA | USA | Astrobotic, IntuitiveMachines, Orbit Beyond. | 2021 | Various; https://www.nasa.gov/press-release/nasa-selects-first-commercial-moon-landing-services-for-artemis-program |
| Planetary Transportation Systems | Germany | ALINA | TBA | Near Apollo 17?; https://ptscientists.com |
| ispace | Japan | HAKUTO-R | 2022 | Near Lacus Mortis pit?; https://ispace-inc.com |
| Moon Express | USA | Lunar Scout | | http://www.moonexpress.com/expeditions/ |
| Astrobotic | USA | Peregrine Griffin | 2021 | Lacus Mortis; then user driven; https://www.astrobotic.com; https://www.nasa.gov/press-release/nasa-selects-first-commercial-moon-landing-services-for-artemis-program |
| Masten | USA | XL-1 | 2021 | TBA; https://www.masten.aero/lunar-vehicles |
| SpaceIL | Israel | Beeresheet | 2019 April 11 failed | Mare Serenitatis; https://www.timesofisrael.com/in-first-israeli-spacecraft-set-for-trip-to-the-moon/ |
| Intuitive Machines | USA | TBA | 2021 | Oceanus Procellarum; https://www.nasa.gov/press-release/nasa-selects-first-commercial-moon-landing-services-for-artemis-program |
| Orbit Beyond | USA | TBA | 2020 | Mare Imbrium; https://www.nasa.gov/press-release/nasa-selects-first-commercial-moon-landing-services-for-artemis-program |

Over the past two decades the lunar surface and sub-surface have been mapped in increasing detail by a succession of lunar-orbiting spacecraft [6]. While it was once thought that "most of the [lunar] resources are evenly distributed" [7], we now know better. The Moon is far from the undifferentiated "magnificent desolation" so strikingly described by Buzz Aldrin during the historic 1969 Apollo 11 moon landing. Some small areas are far more attractive mission destinations than the rest of the lunar surface.

Over the next half decade, at least ~~six~~ five sovereign nations have credible plans to land on the Moon (China, India, Japan, Russia, USA). In addition, several commercial companies (including PTScientists, Moon Express, Astrobotic, Masten, ispace), and the non-profit SpaceIL, have stated intentions to do so. The US National Aeronautics and Space Administration (NASA) plans to use one or more commercial companies to provide transport for its payloads to the lunar surface [8]. Table 1 lists recent and announced soft landing attempts. From this table it can be seen that although the choice of landing sites is diversifying, by moving away from the Apollo landing sites near to the lunar equator. There is also some pooling together in the newer proposed landing sites, e.g. in the South Polar region. Choices have been influenced by the possible presence of the strategic resources detailed in the next section. Even though many of these efforts could be delayed or could fail, either technically (like Beeresheet and Vikram), or financially (as PTScientists nearly did; they were taken over by the Zeitfracht Group and are now Planetary Transportation Systems [9]), there are enough attempts to make it likely that some will succeed. Indeed, at the time of writing, Chang'e 4 has already done so. In the absence of comprehensive failure or cancellation, and within a short period of time, projected landing sites will become actual landing sites with a pattern of preference for some areas over others.

## 2. Concentrated Lunar Resources

A wave of lunar exploration missions since NASA's 1994 *Clementine* mission has significantly improved our overall understanding of lunar resources. Detailed maps, mostly at 1 km – 100 km resolution, of the varying lunar composition, gravity field and temperature have been produced by *Clementine* (launched 1994), *Lunar Prospector* (1998), *Lunar Reconnaissance Orbiter* (*LRO*, 2009 - present), *Chandrayaan-1* (2008 – 2009), *Kaguya* (2009), *Gravity Recovery and Interior Laboratory* (*GRAIL*, 2011 - 2012), and other missions.

Table 2: Typology of Concentrated Lunar Resources.

| Concentrated lunar resources | | | |
|---|---|---|---|
| **Features** | | **Materials** | **Cultural sites** |
| **Topographical features** | **Special locations** | Thorium<br>Uranium<br>Rare Earth Elements (REEs)<br>Helium-3 ($^3$He)<br>Iron | Apollo landing sites<br>Other historical sites |
| Peaks of Eternal Light<br>Cold traps<br>Coldest traps<br>Far-side smooth terrain areas<br>Pits | 33.1E, 0N<br>Sinus Medii<br>Lipskiy Crater | | |

However, the sites on the Moon likely to attract the most attention for future missions are confined to a number of areas on scales of mere kilometers. These account for a tiny fraction of the Moon's total surface area of 38 million sq. km. They are small concentrations within a total area that is equivalent to around 1.5 times the size of North America. These sites of interest can be organized into three main types, each presenting distinct implications for crowding and interference (Table 2). The various sites also present different forms of value to lunar users pursuing distinct types of mission. We discuss their value and implications for the main types of mission currently envisioned to the lunar surface: scientific, human-exploration, and commercial missions [5].

## 2.1 Lunar Features

*The Peaks of Eternal Light* are regions near the lunar poles, which are almost continuously illuminated by the Sun (and so their informal name is somewhat inaccurate), cover only about a few sq. km [10, 11, 12, 13, 14]. From these peaks the Sun is seen to move around the horizon, bobbing up and down slightly over the course of a year. (Unlike the Earth, which is tilted by 23½ degrees to its solar orbit plane, the Moon is tilted by just 1½ degrees; hence there are no seasons on the Moon.) From a sufficiently high peak the horizon is low enough that the Sun is visible even when it is at its lowest point. Local topography limits truly uninterrupted sunlight to a few months at a time [13]. These Peaks are valuable for both the collection of almost continuous solar power, and as locations where the ~300 degrees Celsius day to night temperature swings of the typical equatorial lunar surface location [15] are mostly avoided [14]. These features make the Peaks attractive locations for a wide range of missions, including any requiring a stable power source. Solar panels would have to rotate, or be made roughly cylindrical, but they could collect power at all times (with the obvious exception of any eclipse). The Peaks also allow nearly continuous observation of the Sun e.g. with a radio telescope [3]. Since the Sun is always within a few degrees of the horizon at these Peaks then, if installing a solar power-collecting towers would cast long shadows that could prevent power being collected at other locations, especially if they are tall. Such shadowing is an obvious source of potential disputes. Kilometer-high solar power towers may be quite feasible on the Moon due to the low gravity, lack of atmosphere, and seismic quietue. In this case more areas will provide "eternal light" [14].

*Cold Traps* in the permanently dark craters at the poles are thought to contain volatile materials from the early solar system, including water [16]. The largest of these cold traps are about 50 km in diameter (see figure 1, [17]). The cold traps are the floors of craters whose rims may be Peaks of Eternal Light. However, the traps are likely to cover considerably larger areas than the Peaks of Eternal Light. The long shadows of these rims hit high on the crater walls, and have left the floors in almost total darkness for up to 3.5 billion years [17]. Illuminated only by starlight and reflections off the nearby rims, the traps remain extremely cold (below -180 C, or no more than 90 degrees above absolute zero). These are temperatures where oxygen liquifies. (Not that liquid oxygen would be stable in these cold traps.) Surface ice is stable even in a vacuum below -163 C and is detected [16]. Other water may be present below the surface [18]. Not all cold traps show signs of water [18]. Surface ice appears to be patchy in the cold traps, as ice signatures are only seen in 3.5% of cold traps [16, 19]. Estimates suggest as much as a billion tons of water may be in these traps [20], though this number is highly uncertain [21]. Higher resolution neutron mapping is really needed to obtain an accurate estimate. From one point of view this is a lot, enough to launch the Shuttle over a million times or to supply a city of a million for a thousand years, so long as they adopt recycling as efficiently as on the International Space Station. From another viewpoint, a billion tons is a small amount; the Hoover Dam holds back about 35 times as much water in Lake Mead [22].

*The coldest of the cold traps* reach down to temperatures less than 50 degrees and as low as 25 degrees above absolute zero. They are of order 1 km in diameter (see figure 1, [17]). At such temperatures oxygen is a solid. (Again, not that we expect to find solid oxygen in the coldest cold traps.) If there are volatiles that are only retained at such low temperatures then these coldest traps will

55

be of special interest to scientists and possibly to commercial actors. The coldest traps will also present unique opportunities for a range of other scientific missions. They may be uniquely well-suited sites for far-infrared telescopes, readily keeping the telescope structure so cold that it does not radiate at the same wavelengths (~100 μm) at which astronomers want to observe the sky. They may even help with the building of interferometers using ultra-cold atoms to test fundamental physics and the nature of Dark Matter [23]. While ultra-cold atom physics require much lower micro-degree temperatures [23], most of the power required is needed for the initial cooling from room temperature tens of degrees above absolute zero. The coldest traps just may then be a preferred place to build ultra-cold atom facilities on a far larger scale than on Earth or in laboratories in free space.

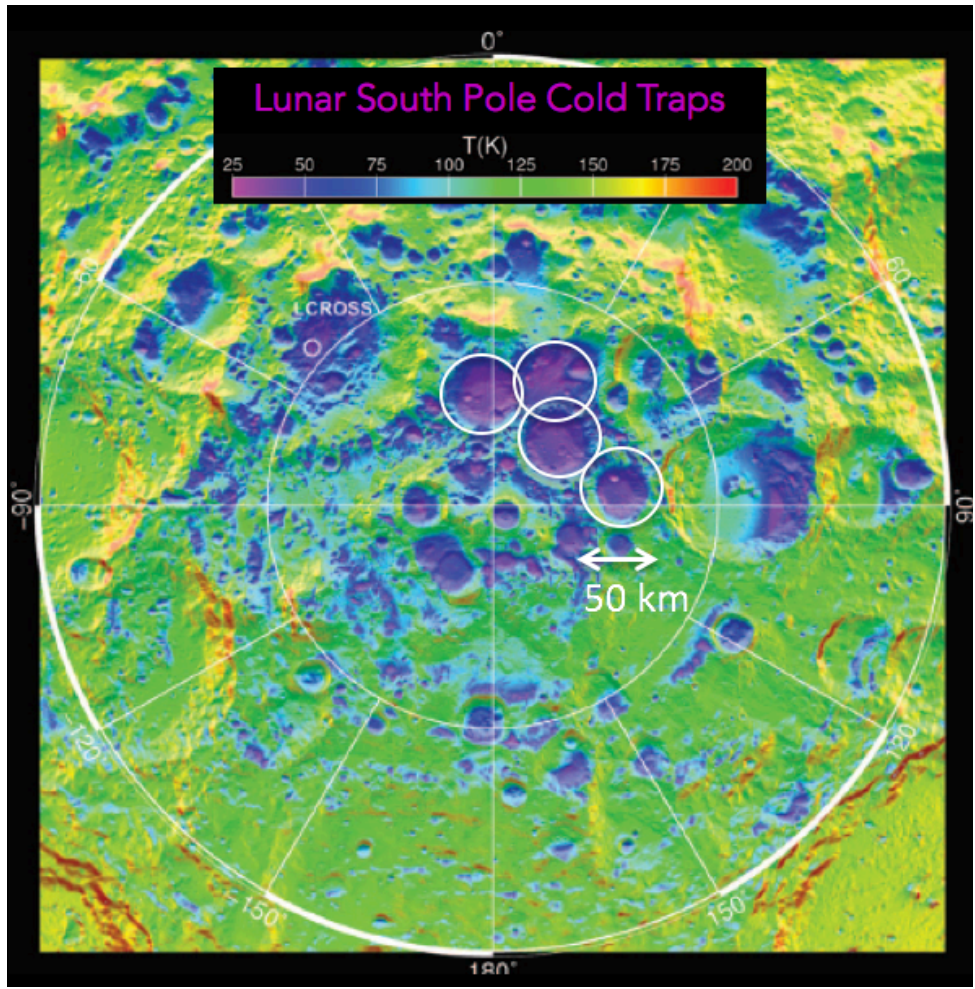

Figure 1. Lunar cold traps at the South pole [18]. The 4 white-circled regions contain the coldest terrain with average annual near-surface temperatures of 25-50 K (i.e. above absolute zero). They are about 50 km across. Reproduced with the permission of David Paige.

*Large Far-side Smooth Terrain Areas* for the emplacement of a cosmology telescope [24, 25, 26]. The lunar far-side is shielded from terrestrial radio emission and so is a natural "radio-quiet zone" for a sensitive cosmology telescope. While there are many location options for initial ~15 km sized radio arrays that can do great astronomy by imaging the largest cosmological structures on a few arcminute scale [27], structures (or multipoles) that are 10 times smaller carry much additional information [28]. To image these smaller structures would require an array some 150 km across. (The resolution of a telescope is given by the ratio of the wavelength being studied to the diameter of the telescope. For the multi-meter wavelengths being studied for cosmology this demands a diameter of some 200 km.) However, the far-side does not have the extensive smooth terrain of the lunar near-side maria. As a result, there are only six areas large enough to incorporate the roughly 200 km diameter area needed for the ultimate resolution (Mare Moscoviensis, Mendeleev, Mare Ingenii, Korolev, Apollo, Hertzsprung). This size requirement comes from the need to provide sufficiently detailed images that structure in the neutral hydrogen at times before any stars had formed can be mapped on fine enough scales to test cosmological models. In addition, such telescopes could distinguish foreground galaxies from the earliest, and most distant, signals from the time before there were stars or galaxies. This goal would translate into a requirement for a resolution of a few arcseconds. With so few suitable locations for the full-size telescope array, conflicts with other potential uses are likely. As an example, figure 2 shows an image of one of the most promising sites, Mare Moscoviensis. Overlain is a map of $^3$He concentration from Kim et al. [29], showing that one of the few far-side concentrations of $^3$He occurs within Moscoviensis. A far-side mining location is likely to cause fewer objections as no mining "scar" will be visible from Earth. If the use of these few locations were pre-empted by other users – virtually all of which are likely to produce electromagnetic interference making them useless for radio astronomy – astronomers would be unable to push to these limits for a long time.

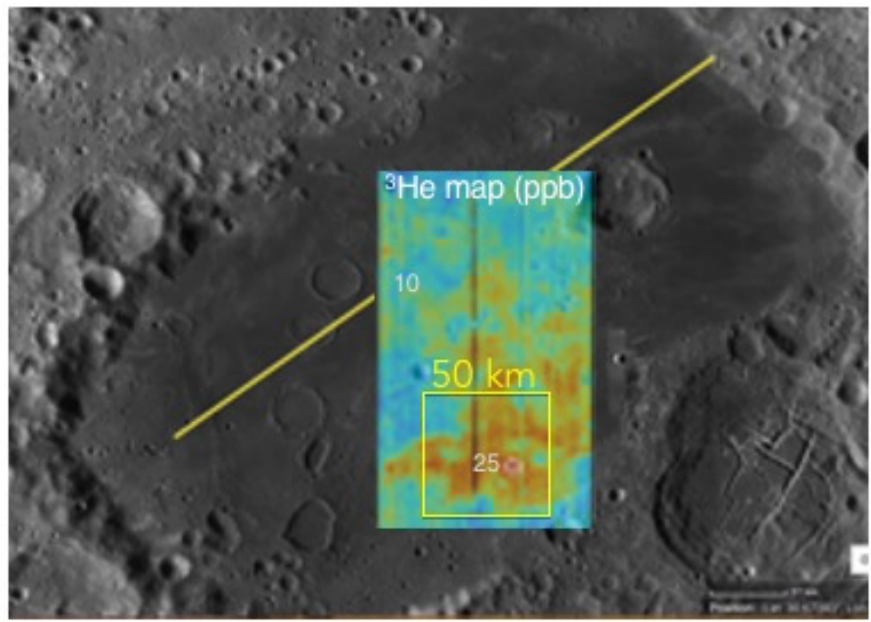

Figure 2. Image of Mare Moscoviensis (grey scale) from the JPL moon trek tool [trek.nasa.gov]. The yellow bar shows a length of 250 km, suitable for a full-scale far-side radio telescope array. The color overlay shows the concentration of $^3$He from Kim et al. (2019) [29]. Concentrations of $^3$He in ppb are given in white. Reproduced with the permission of Kyeong Kim.

*Lunar Pits* potentially offer access to radiation- and meteorite-protected environments that are at a moderate, constant temperature, making them particularly interesting to human-exploration or settlement missions. There are 221 known in the survey of Wagner and Robinson from 2014 [30]. There may be more pits near to the lunar poles, where the almost horizontal illumination makes them hard to spot, although the volcanic flows from which they form are largely absent there. The most valuable pits will be those that have significant overhangs or that lead to intact (uncollapsed) lava tubes that can provide ready-made radiation protection. The evidence does support their presence (e.g. at the Marius Hills in Oceanus Procellarum) [31]. However, they seem to be rare. And only the subset of those pits that are readily accessed can make for good early use. Any of these qualifying pits that also lie near to other resources will be especially valuable and are likely to be few in number.

There are also a number of special locations that may offer advantages for future technologies (mass drivers and space elevators) by providing possible ways to reduce the costs of lifting lunar resources into space [32]. These may close the business case for profitable mining and for space infrastructure construction. Their utilization may be decades in the future but are not inherently infeasible. Although they have no role in current mission planning, their future use cannot simply be discounted. If lunar mining becomes a reality, then they are likely to be given serious consideration, with prospective sites targeted as strategic well before the actual development of the technologies in question.

*33.1E, 0N*, on the lunar equator, just west of crater Maskelyne A, is the optimal place to locate a mass driver to the Earth-Moon L2 point. This precise location is mountainous, but about 20 km distant are two suitably flat areas a few kilometers across [33].

*Sinus Medii* being on the equator at the prime meridian (0E, 0N) is the only location at which a lunar elevator to the Earth-Moon Lagrange L1 point could be sited [32]. It would also be a convenient location to locate an antenna to receive microwave power beamed from a solar power station at the Earth-Moon Lagrange L1 point, should such a technology become viable. Sinus Medii is quite flat and about 287 km in diameter [34].

*Lipskiy Crater* (179.38E, 2.15S) is only ~70 km from the antipode to Sinus Medii and so is a prime location to site a lunar elevator to the Earth-Moon L2 point [32]. It is 91 km in diameter [35]. A smaller flattish region closer to the equator is another candidate.

## 2.2 Lunar Materials

*Thorium- and uranium-rich regions* could in principle be mined for radioactive fuel, though even the "high" concentrations of thorium (~10 ppm) on the Moon are low by terrestrial standards [36]. Thorium and uranium are found together [36]. The highest concentrations lie in 34 regions that are certainly less than 80 km across [36] and may be much smaller. Iron oxide (FeO) is anticorrelated with the presence of thorium and can be mapped in finer detail [36]. In these maps the strongest minima, representing the richest thorium deposits, are only a few kilometers across. The anticorrelation may not be a reliable measure though; improved direct measurements of thorium are really needed.

*Rare Earth Elements* are not actually rare on Earth, but they are not highly concentrated, and their extraction is difficult and highly polluting. Their new-found importance for technology gives a political and strategic value to having reliable supplies. The Moon contains a region of enhanced rare earth element concentrations in the "KREEP" zone of the Oceanus Procellarum [37] (the right eye of the "Man in the Moon"). KREEP stands for "potassium (chemical symbol K), rare earth elements, and phosphorus (chemical symbol P)". This KREEP Terrane province appears to have been among the last regions of the lunar surface to solidify, leading to this unusual concentration. It is not clear that REEs are sufficiently concentrated enough to be ore-bearing [38]. More detailed mapping is needed. KREEP is generally found where thorium is found and so Thorium may be a guide to high REE concentrations [37].

*Helium-3* ($^3$He) is often promoted as a unique lunar resource to fuel fusion reactors, as it is captured by the lunar regolith from the solar wind. (However, we do not yet have fusion reactors and any use of lunar $^3$He is decades away, at best [20].) Although widespread in the Maria at low concentrations (10 - 15 ppb), indirect mapping now shows that there are about eight regions with somewhat higher concentrations of $^3$He up to ~25 ppb [39, 29]. All eight are relatively small (<50 km across). One of these areas lies in Mare Moscoviensis, a promising site for a cosmology telescope. It is notable that the study highlighting these enhanced $^3$He concentration regions [29] was not a purely scientific study, but a prospecting based one, combining $^3$He concentration mapping with LRO terrain data to find the $^3$He-rich sites with easy landing areas.

*Iron-rich regions* derived from asteroid impacts are also quite small (~30 – 300 km across) and are limited to 20 or so sites [40]. These will be quite easy to process being largely metallic and macroscopic. Other sources of lunar iron, though widespread, will be harder to process both energetically and because the iron comes in sub-micron sized grains [20]. Asteroid iron also has the advantage that it may also be rich in precious metals, including platinum and palladium. Iron could be used to construct mass drivers to cheaply send mass to orbit or for settlement infrastructure. Iron is detected by finding lunar magnetic anomalies. Converting the magnetic field strengths to a mass of iron is uncertain at best.

The true sizes of the resource-rich regions may be smaller than we currently can measure. Most of the maps are limited by the spatial resolution of the available instruments. If so, then the true resource concentrations may be higher than now measured and would lie in smaller areas. As an example, figure 3 shows how improving the mapping resolution by a factor three for thorium has revealed that the thorium-rich features are more concentrated than they appeared to be in the lower resolution data [36]. There is a clear need for improved prospecting, particularly in neutron and gamma-ray imaging.

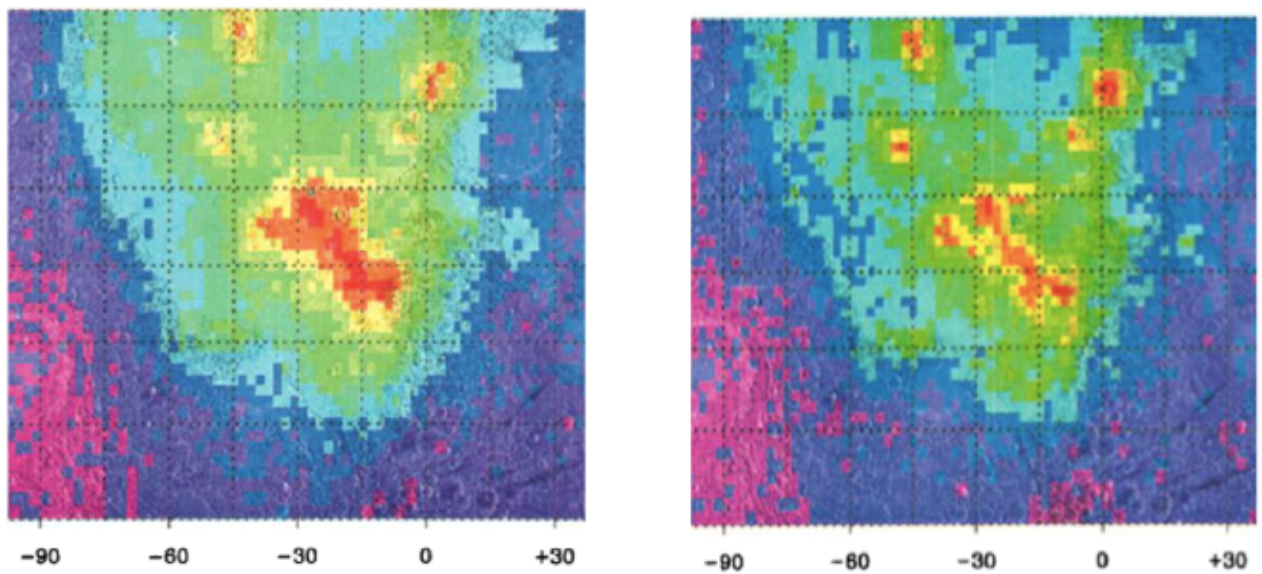

Figure 3. Thorium concentration maps, imaged by Lunar Prospector [36]. The left-hand map used data from 100 km altitude, while the right-hand map has 3 times more resolution (~50 km), being taken from 32 km altitude. The clumping up of thorium rich sites into smaller regions is apparent. Reproduced with the permission of Ben Bussey.

2.3 Cultural Sites

Finally, there are the historical sites, notably the six Apollo mission landing sites. These have some scientific value, but they are primarily "Lunar Heritage Sites" [41], comparable to UNESCO World Heritage Sites. They meet the first of the 10 qualifying UNESCO criteria "*to represent a masterpiece of human creative genius*" and may meet criterion (vi) which is best applied in conjunction with other criteria: they are "*directly or tangibly associated with events…of outstanding universal significance*" [42]. There is also scientific and engineering interest in these sites as they are natural experiments in exposing microbes [43] and manufactured parts to space conditions for over 50 years.

There are other historical sites too, of both hard and soft landings, successful and not. In the longer term, they could even become tourist sites. These sites present not only the question of *how* to responsibly coordinate activities at them, but also, more fundamentally, the question of *whether* we should exploit them in the first place. Meanwhile, some are working to ensure their preservation, notably the non-profit organization For All Moonkind [44]. NASA has developed preservation recommendations for these sites [45]. These recommendations are strictest for Apollo 11 (the first landing), and Apollo 17 (the last landing). For these two sites, minimum approach distances for rovers are 75 m and 225 m respectively, reflecting the larger distances traversed by the astronauts in the later mission. For the other Apollo sites (15, 18, 19) NASA's recommendations are for rover buffer distances of just a few meters from the emplaced hardware, while maintaining 200 m landing exclusion zones. (Apollo 12 landed, by design, just 183 m from Surveyor 3 [46]. A more extensive zone there may be advisable.)

## 3. Disputes over "Potentially Harmful Interference"

If conflicts over lunar resources arise in the coming decade, as seems probable, they will incentivize searches for creative interpretations of the only applicable treaty with broad international recognition, the 1967 Outer Space Treaty (OST) [47]. More specifically, they may invite creative interpretations of Article II's explicit statement that "Outer space, including the Moon and other celestial bodies, is not subject to national appropriation by claim of sovereignty, by means of use or occupation, or by any other means." While the letter and the spirit of the Treaty prohibit formal appropriation, some of its provisions may in fact enable unexpected forms of de facto appropriation. In particular, Article IX introduces the principle of parties' "due regard" for the activities of other parties. The Treaty also states that, if a party's activity could cause "potentially harmful interference with activities of other States," parties can enter in consultations to address the matter. These concepts have enduring relevance. A statement of principles for the Artemis Accords, an architecture of bilateral agreements for lunar cooperation proposed by the United States in 2020, reaffirms commitment to Article IX and emphasizes a duty for parties to coordinate with and notify each in order to prevent interference [48].

These provisions in view, we recognize that parties could invoke their research activities to seek the exclusion from nearby areas of others whose activities present interference risks. At minimum, where significant resources are at stake, it seems likely that disputes over expectations and the practical meaning of "due regard" will arise and require resolution. No mechanism for resolving such disputes currently exists. We argue here that our previous work on the Peaks of Eternal Light [3], identifying the likelihood of competition for this limited resource, is not a special case. Disputes over entitlement to access and entitlements to exclude, in order to prevent "potentially harmful interference," will apply in many cases, independent of the local resources or the lack thereof. But they are especially likely to occur at, or near to, the strategically valuable locations where lunar resources happen to be concentrated.

The small number and size of these sites, coupled with the significant number of missions planned for within this decade, portends crowding and interference between activities at these locations, even while the vast majority of the lunar surface remains untouched. Without raising any issues of overall lunar protection for science, or ethical concerns about lunar integrity [49], the diverse actors targeting strategic resource sites may modify *local* areas of the lunar surface in a range of ways that serve their own purposes but undermine other actors' plans. Even at this relatively early stage, we can already anticipate and prepare for several likely types of interference.

Many scientific sensors are extremely sensitive to electrical signals, light, vibration, dust, and mechanical damage. This sensitivity is both deliberate and necessary, as the signals they are typically seeking are miniscule. Astronomical telescopes are particularly clear cases as they must be open to the sky, and their large mirrors must remain clean in order to carry out their function. As a result, entirely legitimate experiments may require avoidance zones for any other nearby activity. For instance, NASA's statement of principles for the Artemis Accords explicitly provides for "safety zones" [48]. Such avoidance zones might need to be quite large, but their size could in some cases be reduced by installing landing pads to reduce dust and other effects. Given this potential, landing pads could be among the first forms of shared infrastructure that actors jointly build or use near "high-traffic" lunar features.

Two examples, that will apply anywhere on the Moon, illustrate the value of such measures: landing and blast ellipses.

*The Landing Ellipse*. If a spacecraft were to land on another, or sweep its rocket exhaust over it, that would likely damage the already in-place spacecraft beyond use. So, planners of a new lander need to have high confidence that they will not land too close to another. At the moment the uncertainty in the landing point for a lander is a few 100 meters, to have no more than a 1 in 10 chance that it will land somewhere in that zone [50]. Hence the exclusion zone for any lander already emplaced needs to be about a kilometer in diameter. NASA recommends a more stringent 2 km minimum distance (radius) based on a tougher 0.3% chance (3 sigma) of interference [42]. This region will shrink as better navigation technology (e.g. terrain relative navigation [51]) becomes available.

*The Blast Ellipse*. A larger exclusion zone is needed to avoid a new lander causing an already emplaced lander to be hit with rocks and dust blasted up by its rocket plume when landing, or at take-off [52, 53]. This blast of lunar surface material would coat the surfaces of all instruments with dust and may get into mechanical moving parts and make them stick. In contrast to the previous case, this radius cannot be shrunk with new technology as the rocket must always have the sufficient power to slow its descent. As our rocket technology improves, larger and larger payloads will be brought to the surface and their blast ellipses will grow, up to the point at which regolith (loose surface material) will be removed down to larger scale rock that will not be ejected over large distances. This layer is likely to be several meters down [53, 54]. Studies of the Apollo 12 landing find that several tons of regolith were removed by its rocket plume, and that small dust particles achieved escape velocity from the Moon [55]. Slightly larger particles will travel large distances. It is thus impossible to avoid *all* contamination of an emplaced lander by a later lander, however far away it touches down. Determining a safe size for an exclusion zone will then depend on the details of how much dust of what size travels what distance, and on what reasonable mitigation measures the new and the already emplaced landers could have taken. At minimum, estimates calculated from both physics and engineering perspectives should inform how these limits are set, although a broader range of disciplines will need to be drawn upon in drafting recommendations.

There are at least two other cases: A far-infrared telescope such as the OWL-Moon described in these proceedings [56] must necessarily have its mirror open to space to operate. The natural level of dust lofted by meteor impacts [57] may be greatly enhanced by any nearby mining activity to extract valuable water or volatiles must also not kick up dust, as that would coat the mirror and render it far less sensitive. An array of several telescopes could then be used to create quasi-property applying to a whole cold trap, denying others access to its large resources of water. As already noted, not all cold traps contain water [18, 21], so it could be argued that the "dry" ones should be reserved for scientific experiments and telescopes, while the majority of the water-bearing cold traps should be reserved for mining. Similarly, a powerful radio transmitter on a new lander could interfere with instruments on a nearby, emplaced lander (as in our example of the Peaks of Eternal Light [3]). In this case the exclusion zone may be defined by the horizon, which is about 2.5 km away for a typical human eye height (1.7m) [58], but is farther away for objects at higher elevations, particularly on a mountain peak or a crater rim.

Each of these cases would constitute harmful interference. It is important to notice that they also apply on a scale comparable with many of the concentrated resources detailed in section 2.

Exclusion zones for new landers do not preclude the exploitation of a resource within that zone by another party. They only complicate the process. The new lander can deploy a rover to go over to the prime site. In fact, PTScientists planned to do this for their first lunar lander mission. Their rover had only a small fraction of the mass of the lander, and so will carry far more limited equipment. As this will likely be generally true of rovers, the ability to engage in resource utilization by the later arrival will be significantly hampered.

The intention of PTScientists was to land a few kilometers from the Apollo 17 site in the Taurus-Littrow valley. Their rover would then go over to the site and take pictures. PTScientists said that they will treat the site "like a cathedral" [59], and they worked with "For All Moonkind Inc." [44] and with NASA to preserve the site. (The post-reorganization PTS have not said if they are continuing with a landing near to Apollo 17.) Moon Express and Astrobotic have also announced that they will comply with the NASA recommendations regarding lunar heritage sites [41]. The requirement for US Federal Aviation Authority approval for payloads adds pressure to comply [60, 61]. However, there is currently nothing in international law to prevent them, or their less scrupulous successors, from doing as they wish with the site.

Access by rovers could be countered by technologically justifiable means. A user could try to create exclusive access to a small site with valuable resources, by putting up "fences". Under the OST they cannot be literal fences, which have only one purpose – to exclude others. Perhaps, though, they could place the solar panels and batteries that they need in a ring and connect them with cables strung on poles to avoid damage by the surface (figure 4). As it happens, they create a closed loop around the resource that a rover can only breach by causing harmful interference, giving the first arrival de facto exclusive control. In response, a rover with legs instead of wheels might simply step over the cables. The possibility of an "arms race" of such measures and countermeasures is clearly present.

Given these forms of interference, attempts to exclude future actors from lunar features or the effective appropriation of these features appear possible. Partial precedents from comparable environments suggest that actors on the Moon would attempt to establish exclusion zones. For example, the partners in the International Space Station (ISS) assert a 200 meter "keep out zone", [62]. More recently, the private company Bigelow Aerospace in 2013 successfully sought a review by the U.S. Federal Aviation Administration's Office of Commercial Space Transportation of its plans for a lunar habitat, which included a zone of operation that other U.S. entities would be barred from entering [63]. Given these conceptual precedents for exclusion zones, it is plausible that actors modifying or building installations on or around lunar sites of interest would take similar steps to limit others' access to them. Just this year, a NASA statement of principles for the Artemis Accords provides for creating "safety zones" that could entail such restrictions [48]. The potential for disputes is aggravated by the fact that no forum exists for the "appropriate consultations" between parties facing interference called for in Article XII of the OST. Efforts to address such eventualities are more likely to succeed if they are pursued before new *facts on the ground* are established.

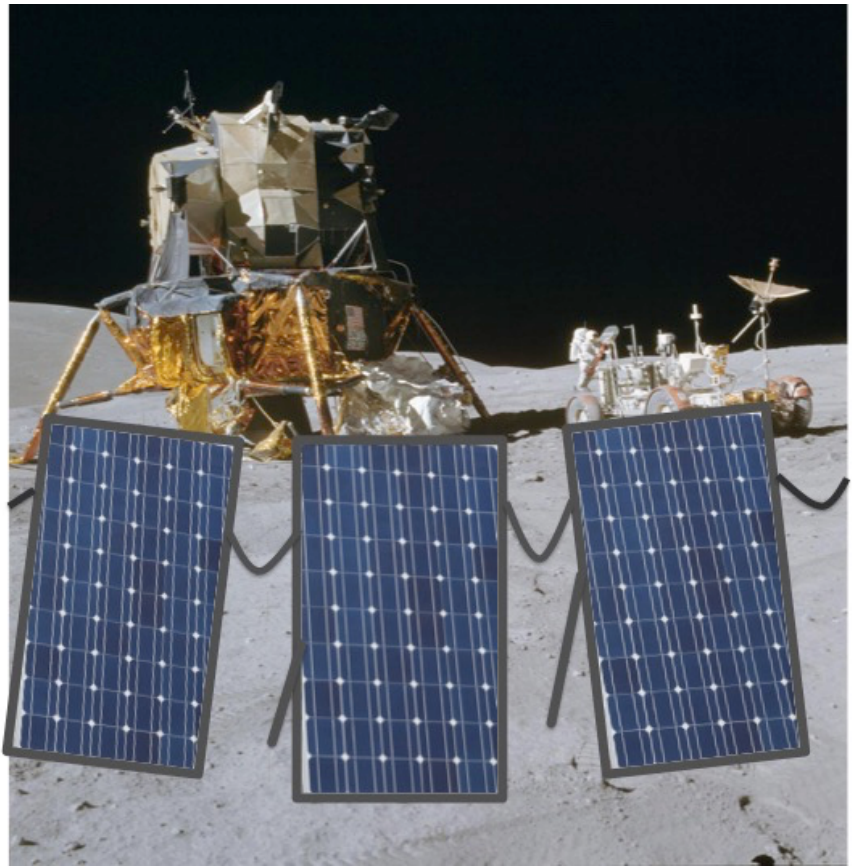

Figure 4. Apollo 16 lunar module and rover surrounded by a (hypothetical) ring of solar panels and cables, thereby restricting access. Modified by the authors from NASA images.

## 4. Policy Considerations

Means of managing these challenges at lunar sites of interest are suggested by coordination and governance mechanisms developed to address similar problems in other settings, most of them terrestrial. From the perspective of policy studies, lunar sites of interest present analogs to global "common-pool resources" or "commons" [64, 65]. These are resources over "which no single nation has a generally recognized exclusive jurisdiction" [66]. In outer space, other comparable environments of interest include orbits and other celestial bodies, such as asteroids and Mars [66]. On Earth, they include the resources of the deep seabed, the water column beyond territorial seas, the electromagnetic frequency spectrum, and Antarctica [66]. Comparisons between terrestrial common-pool resources and space resources can be instructive, even if one rejects the legal or practical status of space resources as "global commons."

A large body of theoretical and empirical work on commons, both global and local, offers insights applicable to crowded lunar sites. Much of this work aims to characterise and respond to "the tragedy of the commons" that results from the mismatch between a collective interest in sustaining a finite shared resource and individuals' interests in maximizing their use of the shared resource, which leads to its overexploitation [67]. This problem is illustrated in the example of a pasture, held in common by a village, that ends up overgrazed. While it is in villagers' *collective* interest to limit their animals' grazing of the pasture to levels that prevent its depletion, it is in every *individual* villager's best interest to allow their animals to graze the pasture as much as possible. With no higher authority to enforce rules that limit overall grazing, the villagers' individually rational choices will likely lead to overgrazing of the common pasture, leaving all worse off.

Responses to the tragedy of the commons have been twofold. The first type has consisted in models for privatizing the commons. For several reasons, including the OST, large-scale formal privatization of lunar sites is not likely in the near term, so we set aside that approach for now [66, 68]. The second type of response has consisted in efforts to theorize and study institutions to manage the problems of overuse and other failures of collective action that arise in the commons, a tradition whose start is credited to Elinor Ostrom [69, 70]. In this section, we distill the findings from this second body of literature relevant to the circumstances of early movers on the Moon, which we define as the twenty or so entities with credible plans to land there before 2030. Scholarship on management of the terrestrial commons presents at least seven general findings that, we suggest, can inform how the early movers approach managing the lunar commons.

## 4.1 Iterate between principle and practice

The management of common resources is most likely to succeed if it emerges from a figurative dialogue between the development of international principles and local experimentation with specific mechanisms. Neither emphasis alone can lead the way.

On the one hand, decades of work on international principles for resource activities demonstrate that developing a common conceptual vocabulary is indispensable for governance in new areas. In this vein, experts focused on space resource management have proposed and given meaning to principles such as "common heritage," "equitable sharing," and "priority rights" since the OST [48, 68, 71, 72, 73, 74, 75, 48].

On the other hand, recent developments highlight the limits of approaches that prioritize achieving broad-based international agreement on principles ahead of attempting practical local steps [76]. For instance, the fate of global climate treaties warns that the pursuit of comprehensive agreement on a framework of principles and subsequent protocols is slow and vulnerable to reversal. In contrast, recent efforts at governance building grounded in user-defined practices or "norms of behavior" have shown promise or, at minimum, appeal to major spacefaring states [for examples, see 77, 78, 79]. NASA's proposed Artemis Accords, expected to consist in bilateral agreements between the United States and other states contributing to the U.S.-led Artemis lunar program, reflect aspects of this trend [48]. These efforts indicate that governance can start with small groups of prospective users focused on immediate solutions to pressing problems, even in the absence of broad international agreement on fundamental principles. Research on terrestrial resources similarly suggests that experimentation and iteration, anchored in co-evolving principles and practices, are integral to the building of effective regimes [80]. By implication, prospects for the governance of lunar resources are brighter if the deliberation of general principles proceeds in tandem with local experimentation in mechanisms for managing interference at particular sites. For example, proposals for an international registry conferring "priority rights" could be prototyped on a small scale by actors at a lunar feature before such a structure is formally established by international agreement [74].

## 4.2 Identify shared interests

Agreement among diverse actors on a desirable long-term outcome for a given lunar site would be conducive to governance, but difficult to achieve. More likely, in the first instance actors will at most agree on what outcomes they seek to *avoid*. Such suboptimal outcomes could include: a scramble for resources that depletes them rapidly and in a wasteful manner; illegitimate or deemed-unfair appropriation of high-value sites by some actors to the exclusion of others; or irreversible damage to culturally significant sites and artifacts [68]. Because actors tend to share an aversion to loss, framing cooperation as loss avoidance can encourage cooperative behavior more effectively than emphasizing long-term gains [81]. Clarifying what is at stake in preventing the least desirable outcomes, through research and deliberation, may help to reframe actors' understandings in a manner that facilitates collective action.

## 4.3 Define the problem

Conceptually prior to the design of any governance framework for a lunar site is the definition of the commons problem it is intended to solve. Problem definition breaks down into at least three distinct aspects: the nature, production, and distribution of the common good [64, 82]. Defining the nature of the commons at stake may entail distinguishing the "resource system" from the flow of "resource units" it generates, or the pastureland from the grass [64]. For example, the Peaks of Eternal Light might be defined as a surface area to be shared, or the solar energy generated by a common facility there might be defined as a resource to be shared. In contrast, defining the production aspect entails specifying the optimal level of exploitation of (or investment in) the common good [82]. Agreement on this level is not straightforward [82], but even if it is achieved, actors will often be tempted to defect from the deal. Finally, defining the distribution scheme entails specifying how benefits from the exploitation of (or investment in) the good will be allocated across actors [82]. Defining a distribution scheme is a negotiated process that is likely to feed into deliberation of production-level questions [66, 82].

## 4.4 Lengthen the time horizon

Cooperation between actors is more likely when the future casts a long shadow, meaning that actors expect to interact repeatedly [83; for qualifications see 84, 85]. Different framings of the same issue can also change how heavily actors discount the future and, therefore, how they decide to act in the present [81, 85]. Early movers may enhance the prospects for cooperation by deliberately promulgating framings of lunar governance that orient participants' reasoning past short-term considerations and toward long-term outcomes. Helpful steps could include defining a common long-term agenda for lunar activity and reducing the scope of uncertainty about anticipated future technological change through discussion and analysis.

## 4.5 Design accommodating platforms

The management of crowding and interference at lunar sites appears most likely to succeed if it is performed through user-created institutions. Proposals for such arrangements on the Moon have already been advanced. Examples include Karen Cramer's proposed Lunar Users Union [68]. Promising institutions for such settings take the form of nested platforms designed for multiple users with diverse goals [64, 86, 87]. Devised for the Moon, such an arrangement would allow for the embedding of local governance mechanisms, designed to manage each specific site, into larger regional groupings or lunar-level structures. These higher-level structures, in turn, would establish common governance principles to guide management across the local sites, enhancing a degree of procedural and conceptual consistency across the various sites without prescribing specific measures for any one of them. Like other "interorganizational systems," lunar governance structures will be more effective if they are designed to accommodate the heterogeneity of participating actors from the start [88, 64, 65]. Moreover, these governance arrangements are most likely to last if they are designed, from the beginning, to evolve with changing circumstances, new users, and new technologies.

## 4.6 Establish habits of cooperation

Historical considerations "play an extremely important part" in accounting for successful outcomes in the management of terrestrial commons [82]. Traditions of collective action can increase its likelihood in the future, suggesting that "cooperation can be habit-forming" [82]. Early movers' choices about whether and how to cooperate on governing could have lasting and disproportionate consequences. If the next wave of lunar missions encounters interference challenges, actors' responses will set precedents that guide expectations and define norms of conduct in response to interference for later missions.

## 4.7 Create withholdable carrots

Actors who are parties to a governance scheme that proscribes or constrains behavior are likely, sooner or later, to face the temptation to defect from the arrangement. The scheme is therefore more likely to succeed if it establishes predetermined consequences to defection or non-compliance. One possible means to fostering this accountability is the creation of new public goods to which actors' access is made conditional upon their adherence to agreements. Examples include common infrastructure, such as waste-management, communications, or power-generation systems and dust-mitigating landing pads [68]. Offenses punishable by denial of full or optimal access to these public goods might include extracting material in excess of an agreed level. The goal would be to deter defection by making it at least as costly as compliance. To be legitimate and equitable, such an enforcement system would need a mechanism through which actors could appeal an adverse decision and should provide other channels through which actors could seek to reform the arrangements or express dissent.

In sum, extensive experience with managing concentrated common resources on Earth suggests lessons for how to manage such resources on the Moon. Effective management of common resources is, in any setting, a complex process, often requiring actors to experiment with imperfect institutional forms while trying to entrench habits of cooperation. The first steps to developing governance solutions consist in characterizing the actors' common and competing interests, defining the problems they face, and bringing long-term outcomes into actors' common view. With a working understanding of these aspects shared among them, actors are more likley to succeed at creating the accommodating institutions needed to deliberate and create management rules for the resource. Governance is likely to be more effective if actors establish cooperative behavior as normal early on in this process and if they create common assets that can later be withheld to deter violations. All of these steps toward creating governance for the commons are best attempted before actors have begun significant activity on the lunar surface or investment in specific mission designs, for reasons both practical and normative, as we explore in the next section.

## 5. Issues of Timeliness and Justice

In a multi-player environment, such as is imminent on the Moon (Table 1), and where there is competition for limited resources, arriving at an effective solution requires building a framework that can be a basis for consensus, or at least for widespread acceptance. As in 4.1 above, this is not the same as arriving at an agreement that everyone actually prefers or regards as optimal. Rather, it is a matter of arriving at an agreement that enough of the major parties can recognize as reasonably just and can live with, irrespective of concessions made. Without such agreement, any framework is liable to lack stability and to break down. We take it that stability is a desirable feature of any account of governance in space, perhaps even a necessary feature. Stability need not be "forever" but should be "for long enough" to limit the risks of the most damaging outcomes [89]. Such damaging outcomes may be thought of in terms of lunar development and lunar protection, both of which could go well or badly.

For stability to be secured, standards of justice are required. Actors will not look favorably upon arrangements that disadvantage them in ways that they regard as clearly unfair. This is where policy and ethics meet, and where our argument draws upon considerations of a slightly more abstract sort in order to support the claim that deliberation about is timely. It falls into the "goldilocks zone" for deliberation. This is the period during which, just outcomes are more likely to be secured. Too early, and an approach is liable to be under-informed; too late and (as indicated above) patterns of behavior become almost impossible to change. The right time to deliberate about these matters, from an ethical point of view, is when we know enough, but not everything that we will eventually need to know, about lunar resources for lunar development to occur in a stable way.

Several factors combine to suggest that we are already in this zone. As we outlined in Section 2, our current knowledge of lunar features and current mission plans, while still in development, is advanced enough to suggest the likelihood of crowding problems, to suggest their overall character, and to identify where they are most likely to occur. However, we are still in some respects 'epistemically disadvantaged' in that we lack the knowledge of the Moon that those who develop lunar resources will eventually have. Yet the limitation of our knowledge also has a clear advantage in two respects:

(a) We do not need to deal with the histories of habitation and use by humans, which are inseparable from detailed *in situ* knowledge. Certain kinds of knowledge come with history, and a more complex set of requirements if justice is to be at all possible. At present, deliberations can remain streamlined because no one can claim any special relation to the Moon, or to strategic resources that others lack. The only special claims relate to the relation between launch states, artefacts, landing sites and crash sites. These are important, but far more limited than the claims of special entitlement, which will rapidly emerge under conditions of lunar development.

(b) While self-interest and national interest may be expected to feed into any deliberations, whenever they occur, the actual results of such deliberations may be less easily skewed if the parties involved have only a provisional idea of which policies will tend to maximize their own position. Even with regard to the strategic resources identified above, some of these resources may eventually be downgraded or even bypassed during the course of actual lunar development. Others, which currently appear less important to us, may play a surprising role. Lack of advance knowledge about these matters paves the way to agreement when more precise knowledge could easily pose an obstacle. The Outer Space Treaty is itself an example of this process at work: nations agreed to a deal (which they have not done at any time since) not because they were aware of what would happen, but because they were unsure about outcomes, and eager to avoid future harms. The Treaty was driven as much by uncertainty as it was by predictable outcomes.

Because we can at least provisionally identify the most likely strategic resources, but not the ways that they will be incorporated in any actual system of lunar development, we are behind a genuine "veil of ignorance" [90, 91]. This is a state which is often desired in terrestrial accounts of justice, but it is also a place where we can never truly be with regard to terrestrial resources because we already have a clear idea of how terrestrial resource utilization operates [90]. In the absence of such detailed knowledge about lunar conditions, interested actors may be more motivated to tackle the risks of crowding and interference in reasonably just, opportunity sharing, ways in order to safeguard future opportunities for themselves. In doing so, they will be more likely to arrive at a stable framework for governance, one which can last "long enough" to protect future opportunities.

By contrast, a failure to proactively address the crowding and interference risks prior to the emergence of actual patterns of use and behaviour, increases the risk that the tragedy of the commons will make itself felt: actors will tend to act against shared interests, and against their own best long-term interests, out of a concern that if they do not act in unfair ways now, others certainly will, while they are left behind. Without timely efforts geared towards resource management and site protection, all but the earliest missions are likely to suffer avoidable losses of opportunity. Such losses include the destruction or degradation of the lunar environment, damage to historical sites, and loss of opportunities for scientific research, as well as the compromising of opportunities for a more stable, longer term, pattern of commercial development.

## 6. Conclusions

Many of the useful and valuable resources on the Moon are concentrated into a modest number (tens) of quite small regions (in the order of a few km). Over the next decade, forms of interference and related disputes and conflicts over these concentrated resources may arise, as many actors, sovereign, philanthropic, and commercial, descend onto just a handful of small, high-value sites on the lunar surface. Responsibly coordinating these diverse actors' activities requires recognizing and accommodating their distinct interests and purposes. Any proposed governance arrangement may have to contend with irreducible practical and conceptual tensions between different actors' designs: scientific, commercial, and human-exploration activities may often be incompatible with each other. Moreover, it is likely that these varied actors' plans are best served by different governance arrangements [5].

While this situation presents likely challenges, humankind's extensive experience with managing common resources on Earth suggests that some of these obstacles can be overcome, while the impact of others may be contained with the help of effective institutions. The study of commons on Earth suggests lessons applicable to efforts at governing lunar sites of interest. To start, a promising pathway to building governance lies in the iterative pursuit of both guiding principles by the international community and local experiments with site-specific institutional forms by lunar users. Moreover, to be effective at any scale, governance arrangements must be underpinned by actors' shared understandings of the nature of the resource itself, their varying interests in it, and the problems its shared use is likely to present. This shared understanding is not straightforward to achieve, as lunar resources and features lend themselves to multiple interpretations and valuations. Deliberation and additional research can help in this respect. Governance institutions are also more likely to succeed if they direct actors' attention to the worst-case long-term outcomes that they may seek to avoid, in addition to the benefits they hope to reap. Efforts at governance are more likely to be effective if they take place within accommodating, flexible institutions that allow diverse and evolving actors to participate in creating the rules for use of the lunar sites. Finally, governance mechanisms have a greater chance of averting worst-case outcomes if actors establish, by precedent, expectations of cooperative behavior and consider devising common "carrots" that they can withhold to deter non-compliance.

Now is an appropriate time to begin developing a governance framework guided by these lessons from Earth. Efforts at managing forthcoming disputes are most likely to succeed if they are undertaken *before* vested interests gain too firm a foothold. We are in several respects better placed to erect a just and effective framework for lunar resources at this time than we will be during actual lunar development. We know enough to show that there is a need for a framework, but not so much that the "veil of ignorance" concerning resources is entirely lifted and players are more fully aware of how they may benefit from any particular choice of framework. As solutions that are globally seen to be just are also most likely to be robust and lasting, a framework developed now has a better chance at legitimacy and stability than one shaped by stakeholders who have an established presence, but strongly diverging goals. Lessons from the management and mismanagement of terrestrial commons suggest that action should be taken now rather than later, or at least now as well as later, to develop the governance structures needed to prevent (and later on contain) avoidable and undesirable problems of crowding and interference.

The need for action is perhaps most acutely felt by the astronomy community. The interference inflicted on terrestrial telescopes by large new satellite constellations presents a cautionary tale. Astronomers have found their capacity to shape policy weakened in the face of ambitious commercial projects that threaten both ordinary people's view of the night sky and humankind's quest to understand its place in the universe. Rather than proactive stewards of these public goods, astronomers risk becoming passive bearers of the consequences of choices made by commercial interests and government regulators without their meaningful input. The result could lead to a substantial loss of observations from telescopes, astronomers' most valuable assets, in the present and a many-times-greater loss of scientific opportunity in the future. The lesson for astronomers to heed is that they have important equities to defend in outer space.

If astronomers do not take the initiative to identify and raise awareness of the scientific and public interest in protecting unique lunar features now, they may find themselves unable to do so once these features are under threat from interference and crowding. In this respect, astronomers find common cause with other scientists, such as astrobiologists, and other researchers for whom planetary protection measures are essential. The scientific community today faces both an opportunity and a responsibility to help guard precious lunar sites from the irreversible damage threatened by crowding and interference.

Acknowledgements
We thank David Paige for permission to reproduce figure 1, and Ben Bussey for permission to reproduce figure 2 and Kyeong Kim for permission to include their image in figure 3. An anonymous referee provided many careful comments that greatly improved the paper. We thank Academic Ventures at the Radcliffe Institute for Advanced Study of Harvard University for hosting all of us for a workshop on space resources. ME thanks the Aspen Center for Physics, funded by NSF grant #1066293, for their hospitality when this paper was initiated. AK thanks the University of Missouri Research Board for a grant to support the project entitled *Aircraft, Spacecraft, and Statecraft*, which made research for this article possible.

**Funding Statement**
The TM contribution to the publication was supported by King's College London, with an International Collaborations grant connected to the *Cosmological Visionaries* project. ME thanks the Aspen Center for Physics, funded by NSF grant #1066293, for their hospitality when this paper was initiated. AK's contribution was enabled by a 2018 University of Missouri Research Board grant for the *Aircraft, Spacecraft, and Statecraft* project.

**Data Accessibility**
No primary data is reported in this paper.

**Competing Interests**
The authors have no competing interests.

**Authors' Contributions**
ME provided the initial conception and the analysis of the distribution of lunar resources, the list of imminent landings, and some legal discussion; AK conducted research for the policy section and some other policy-related issues and drafted and revised the parts of the manuscript that pertain to policy and conclusions; TM focused upon timeliness and justice, and revised some of the overall structure for cohesion and inclusions. All the authors engaged in close discussion of the manuscript.